# SCANDIUM ALUMINUM NITRIDE OVERMODED BULK ACOUSTIC RESONATORS FOR FUTURE WIRELESS COMMUNICATION

*Walter Gubinelli, Pietro Simeoni, Ryan Tetro, Luca Colombo, and Matteo Rinaldi*
Institute for NanoSystems Innovation (NanoSI), Northeastern University, Boston, MA, USA

## ABSTRACT

This work reports on the modeling, fabrication, and experimental characterization of a 13 GHz 30% Scandium-doped Aluminum Nitride (ScAlN) Overmoded Bulk Acoustic Resonator (OBAR) for high-frequency Radio Frequency (RF) applications, notably in 5G technology and beyond. The Finite Element Analysis (FEA) optimization process targets the top and bottom metal electrode thicknesses, balancing the electromechanical coupling coefficient and acoustic energy distribution to enhance device Figure of Merit (FOM). Experimental results on fabricated devices employing platinum and aluminum as bottom and top electrode, respectively, demonstrate a quality factor at resonance ($Q_s$) of 210 and a coupling coefficient ($k_t^2$) of 5.2% at 13.3 GHz for the second bulk thickness overtone, effectively validating the simulation framework and hinting at the possible implementation of OBARs for advanced RF filters in 5G networks.

## INTRODUCTION

The rapid adoption of the 5G technology is marking a major shift in our interaction with digital networks [1] [2]. This growth is fueled by the extensive roll-out of 5G infrastructure, which boosts applications across various sectors, including smart cities, telemedicine, precision agriculture, and industrial operations [3]. These enhancements significantly improve operational efficiency, productivity, and the overall user experience by leveraging advanced radio frequency front-end technologies that enhance data transmission capabilities and network reliability. [4]. However, expanding to higher frequencies presents several challenges, notably the need for improved filtering hardware in Radio Frequency Front-End (RFFE) modules [5].

Among the key technological advancements in wireless communication, the development and commercialization of piezoelectric microacoustic RF filters played a crucial role for the deployment of efficient and reliable higher frequencies protocols, such as 3G and 4G [6]. Notable among these are Surface Acoustic Wave (SAW) resonators [7] [8] and Thin Film Bulk Acoustic Resonators (FBARs) [9], each presenting their set of unique challenges. Although foundational for many RF applications, SAW resonators struggle with limited power handling and increased acoustic and electrical losses at higher frequencies [10]. On the other hand, FBARs present increased fabrication complexity and reduced quality factor ($Q$) due to their strong dependency on piezoelectric film thickness. Recently, Polarized Piezoelectric Films (P3F) [11] and Overmoded Bulk Acoustic Resonators (OBARs) [12] have emerged as an appealing alternative to traditional FBARs due to their higher operational frequencies and reduced substrate losses, while maintaining a well-established fabrication process, thus gaining commercial interest [13].

The development of advanced piezoelectric material, particularly highly doped Scandium-doped Aluminum Nitride (ScAlN), offers further enhancements for designing new device platforms and communication paradigms. Due to its enhanced piezoelectric coefficient, increased dielectric constant, and reduced mechanical compliance, coupled with its low dielectric losses and fabrication compatibility with Complementary Metal-Oxide-Semiconductor (CMOS) technology, ScAlN is rapidly establishing itself as a state-of-the-art material in many commercial applications, including Radio Frequency (RF) filters for 5G [14].

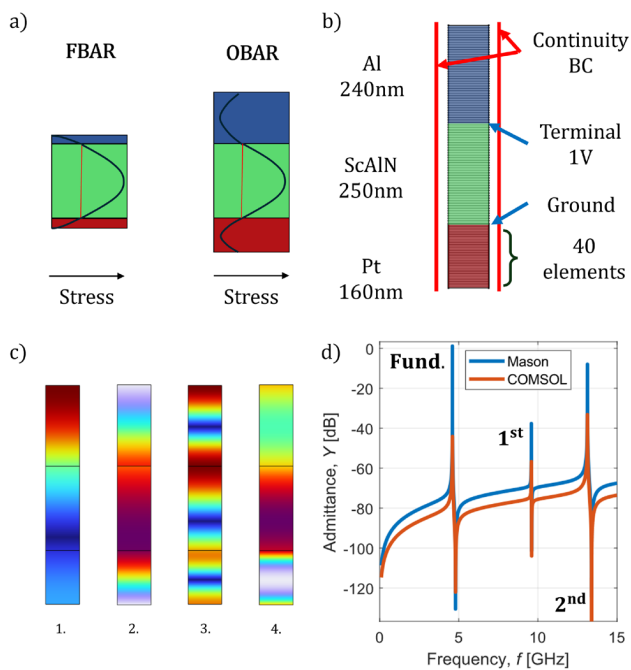

Fig. 1. a) Differences in the stress distributions in the thickness direction ($\sigma_z$) for an FBAR and an OBAR structural element. Each layer is designed to confine half of the acoustic wavelength ($\lambda$) of the desired mode of vibration. b) COMSOL® optimized structural element model. c) Displacement (1 and 3) and stress (2 and 4) distributions for the fundamental tone (1 and 2) and second overtone (3 and 4). d) COMSOL® and Mason model admittance comparison of the targeted modes of vibration. The vertical offset is purposely achieved by setting different out-of-plane widths for a better data visualization. The fundamental mode (4.9 GHz) and the second overtone (13.12 GHz) exhibit an electromechanical coupling ($k_t^2$) of 9.8% and 4.8%, respectively.

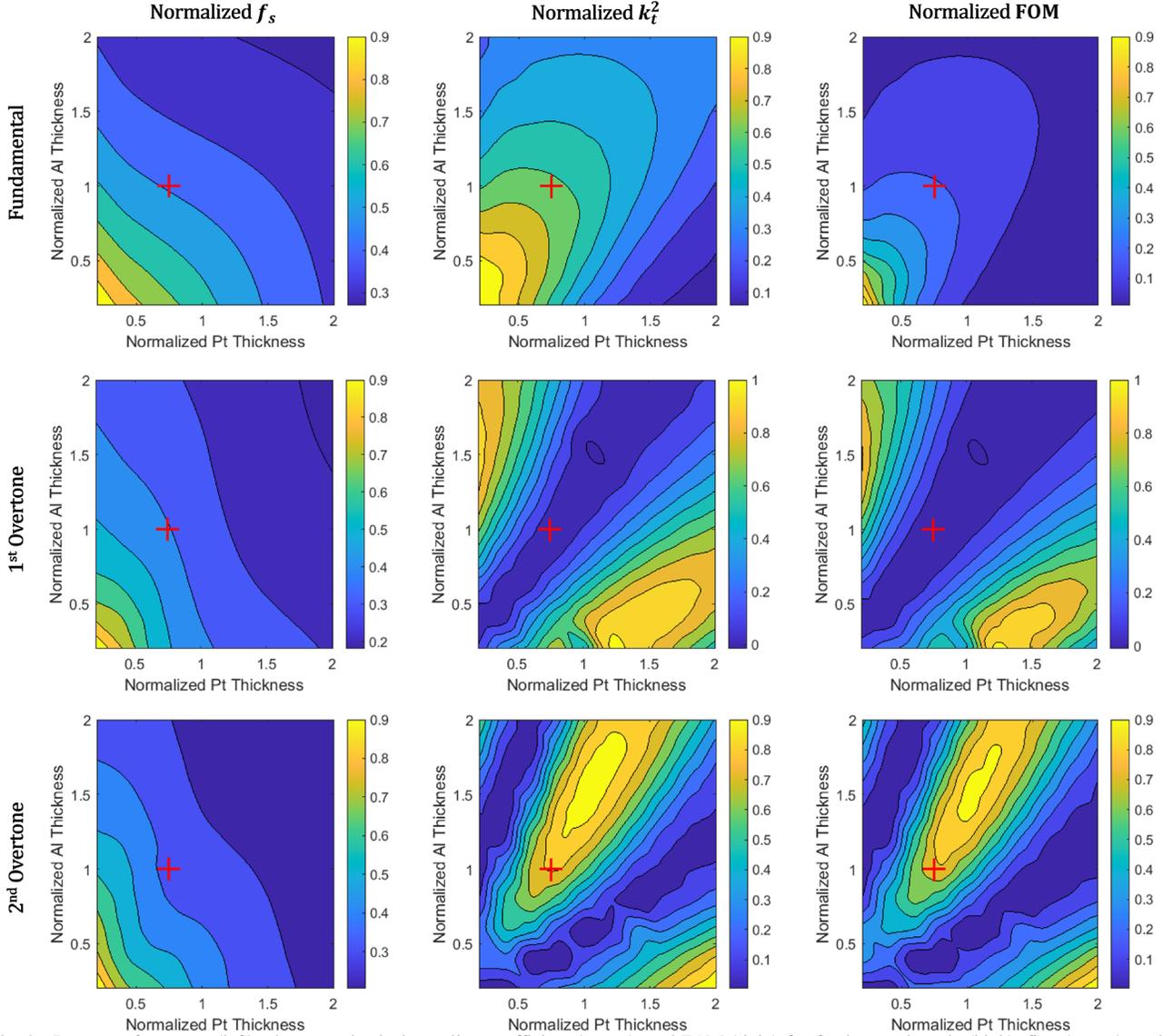

Fig. 2. Resonant frequency (left), electromechanical coupling coefficient (center), and FOM (right) for fundamental mode (high), first (center), and second (bottom) overtone. The resonant frequency is normalized with respect to the natural frequency of the piezoelectric layer, whereas both $k_t^2$ and *FOM* are normalized to their respective maxima.

In this work, the following sections detail modeling, optimization, fabrication, and characterization of a 30% Sc-doped OBAR targeting the proposed mid FR-3 band at 13 GHz. The initial section is devoted to device modeling, focusing on stacking thickness optimization through COMSOL® Multiphysics Finite Element Analysis (FEA) and Mason modeling. The subsequent section outlines the fabrication processes and the experimental validation of the simulation results. Finally, considerations on how to further improve key performance indicators are provided in the conclusions.

## MODELING

OBARs, contrary to FBARs, operate in an overtone thickness mode, divided between the piezoelectric layer and the active metal layers (Fig. 1a). By splitting the acoustic load between the piezoelectric layers and the electrodes, instead of confining the acoustic wave only in the piezoelectric layer, it is possible to achieve higher resonant frequencies while retaining large coupling coefficients. The device under investigation is realized with 30%-doped ScAlN, employing Platinum (Pt) and Aluminum-Silicon-Copper (AlSiCu) as bottom and top electrodes, respectively. Pt is selected due to its compatibility with the growth of highly oriented *c*-axis ScAlN wurtzite crystals [15], while AlSiCu is employed for its low resistivity and density. A COMSOL® quasi-1D model (Fig. 1b) is implemented to evaluate the effect of the electrode thickness on using a 250 nm thick ScAlN layer. With the top and bottom electrode metal thicknesses set to 160 nm and 240 nm, respectively, both values appear to be sub-optimal according to the heat maps. However, they closely match the frequency target while reducing acoustic losses and series resistance. The displacement and stress distribution for both modes of the selected stack are shown in Fig. 1c, while a Mason model is implemented in Advanced Design System (ADS®) to confirm the COMSOL® predictions. The theoretical admittance is shown in Fig. 1d, highlighting a series frequency of 13.15 GHz with a 4.8% coupling for the second overtone. It is worthwhile noticing that each resonant mode rquires a unique stacking combination, as shown in Fig. 2.

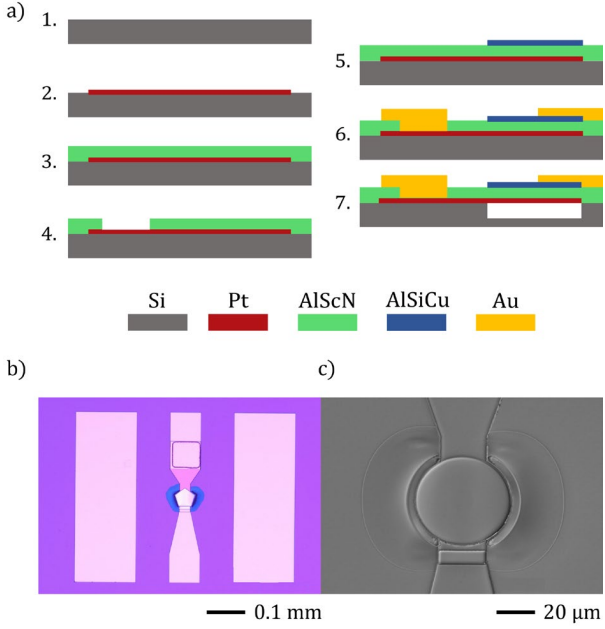

Fig. 3. a) Fabrication steps for the OBARs. 1) HR Si Wafer; 2) Pt patterning and deposition; 3) ScAlN deposition; 4) wet etch vias opening; 5) top electrode patterning and deposition; 6) gold pads patterning and deposition; 7) structure release with XeF$_2$. b) Optical image of device R2C4 (diameter 30 μm). c) SEM-like image of device R5C7 (diameter 35 μm).

## EXPERIMENTAL VALIDATION

A fabrication test run is performed with the identified values for the Pt-ScAlN-AlSiCu stacking to validate the simulation framework described in the previous section. The devices are designed with circular and regular pentagonal geometries, with diameters ranging from 5 μm to 50 μm to evaluate the impact of irregular boundaries on the suppression of unwanted transverse spurious modes due to internal reflection. The plates are tethered to the substrate via symmetrical anchors with a width equal to the radius. The devices are fabricated using a standard FBAR micro-fabrication process consisting of five lithographic steps, as highlighted in Fig. 3. Both electrodes and pads are deposited via RF sputtering and liftoff, employing Titanium (Ti) as an adhesion promoter. The ScAlN layer is the OBAR performance. Several parameters must be considered to optimize the thickness of the metal electrodes, particularly the electromechanical coupling coefficient $k_t^2$ and the energy distribution ratio $\eta$ across the layers. The $\eta$ function, defined as the ratio between the elastic strain energy in the piezoelectric layer and the total strain energy in the stack, estimates the impact of acoustic losses in each domain on the mechanical quality factor at resonance ($Q_m$), according to [16]:

$$Q_m = \frac{1}{\eta \cdot Q_{piezo}^{-1} + (1-\eta) \cdot Q_{metal}^{-1}} \quad (1)$$

where $Q_{piezo}$ and $Q_{metal}$ represent the acoustic losses in the piezoelectric and in the electrodes' domains, respectively [17]. In the model, top and bottom metal thicknesses are swept between $t_m = 0.2 \cdot t_{piezo}$ and $t_m = 2 \cdot t_{piezo}$, and the quality factors for the metal and the piezoelectric layers are set equal to 200 and 2000, respectively. The resulting heat maps detailing series resonant frequency, electromechanical coupling, and Figure of Merit (FOM), defined as $FOM = k_t^2 \cdot Q_m$, as a function of top and bottom metal thickness are shown in Fig. 2. The central frequency of an FBAR resonator is identified by the phase velocity, thickness, and mode order, according to:

$$f_n = \frac{n v^D}{2t} \quad (2)$$

where $n$ is the mode order, $v^D$ the phase velocity in the medium, and $t$ the thickness of the plate. Therefore, a frequency of 13 GHz at the second overtone can be achieved deposited starting from a nominal concentration 12" casted target via RF reactive magnetron sputtering on high resistivity silicon (HR Si). Vias are opened using wet etching in 80° hot phosphoric acid ($H_3PO_4$). Finally, etching pits are opened via ion-milling, and the structures are released with Xenon Difluoride ($XeF_2$) selective isotropic removal of the underlying Si substrate.

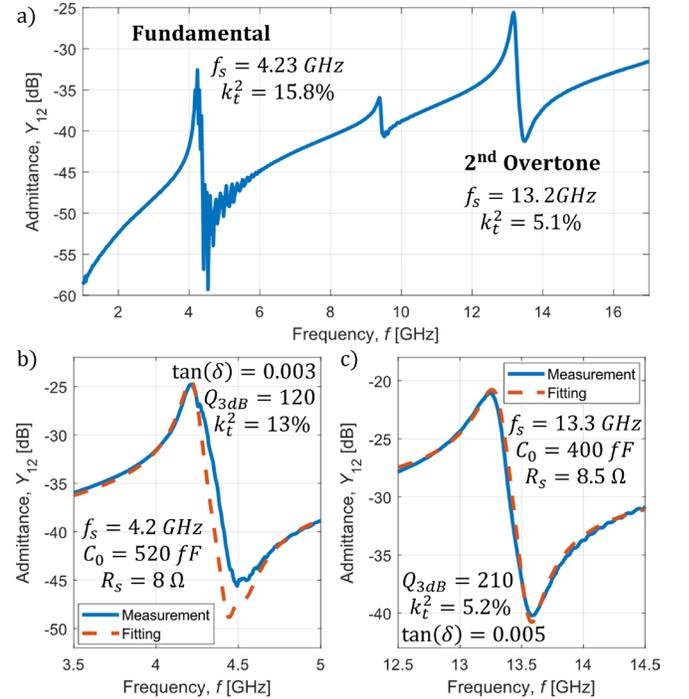

Fig. 4. a) Device R2C4 (circular, 30 μm) broadband response with spurious modes at the fundamental tone anti resonance. b) and c) Device R5C7 (pentagonal, 35 μm) and Device R14C5 (pentagonal, 25 μm) measured $Y_{12}$ admittance response and relative mBVD fitting.

An optical microscope image of a fabricated pentagonal OBAR is reported in Fig. 3b, and a Scanning Electron Microscope image of a fabricated circular OBAR is reported in Fig. 3c. The fabricated devices are later characterized via direct RF probing in a 2-port configuration in laboratory conditions. The S-parameters are recorded with a Vector Network Analyzer after on-chip calibration. The scattering parameters are then converted into admittance parameters via MATLAB®, and the resulting admittance response in transmission ($Y_{12}$) used to extract the motional parameters via mBVD modeling fitting [18]. Compared to the FEA simulations (Fig. 1d), the measured resonant frequencies and $k_t^2$ performance do not precisely match the expected ones. The discrepancies can be attributed different electrode thickness compared to the nominal stacking and top electrode lithography misalignment (Fig. 3b). As expected, circular OBARs present more spurious modes in the fundamental tone due to internal reflections when compared to pentagonal devices (Fig. 4a).

The best fundamental mode is recorded in device R5C7 (pentagonal, diameter of 35 μm), whereas device R14C5 (pentagonal, 25 μm diameter) reports the best overtone performance, as shown in Fig. 4b and Fig. 4c, respectively. The responses of all measured devices are consistent, showing average $FOM = Q_m \cdot k^2 = 10$ for both tones. The higher-than-expected coupling coefficients at the overtone are most likely due to the lower quality factors, which is known to induce a shift in the resonant frequency, leading to an apparent $k_t^2$ increase [19].

## CONCLUSIONS

This work reports on a complete simulation framework for the optimization of FBARs and OBARs stackings, modeling the effects of the metal electrodes on the expected device performances. Quality factors at resonance as high a 210 and $k_t^2$ of 5.2% around 13 GHz are achieved for the second overtone, validating the proposed methodology.

While leaving room for improvement in the fabrication process, the devices show overall good and repeatable performance. Additionally, the $\eta$ model for the strain energy confinement could be improved by introducing a non-constant value for the material quality factor and other non-idealities to appropriately map acoustic losses at higher frequencies and identify electrodes materials suitable for quality factor optimization. To conclude, the above mentioned experimental results validate the proposed models, proving that OBARs are an appealing candidate for high frequency acoustic filters.